\pdfoutput=1
\documentclass[twocolumn,showpacs,preprintnumbers,floatfix,prd]{revtex4-1}
\usepackage{amsmath}
\usepackage{amssymb}
\usepackage{amsfonts}
\usepackage{graphicx} 
\usepackage[bbgreekl]{MyBbol}
\usepackage{mathrsfs}
\usepackage[center,footnotesize,hang]{subfigure}
\usepackage{cancel}
\usepackage{ulem}
\usepackage{hyperref}
\usepackage{color}

\newcommand{\CiteSeeSaw}{\cite{Minkowski:1977sc,Yanagida:1980,Glashow:1979vf,Gell-Mann:1980vs,Mohapatra:1980ia}}

\newcommand{\PL}{\ensuremath{\mathrm{P_L}}}
\newcommand{\PR}{\ensuremath{\mathrm{P_R}}}
\newcommand{\eV}{\ensuremath{\,\mathrm{eV}}}

\newcommand{\GeV}{\ensuremath{\,\mathrm{GeV}}}

\newcommand{\dd}{\ensuremath{\mathrm{d}}}

\newcommand{\Ord}[1]{\ensuremath{\mathcal{O}(#1)}}
\newcommand{\braket}[1]{\ensuremath{\left<#1\right>}}

\newcommand{\Higgs}{\ensuremath{\phi}}
\newcommand{\fmslash}[1]{\cancel{#1}}

\newcommand{\SuperField}[1]{\bbsymbol{#1}}

\def\D{\mathrm{d}} 
\def\I{\mathrm{i}}
\def\SU{\mathrm{SU}}
\def\U{\mathrm{U}}

\DeclareMathOperator{\re}{Re}

\DeclareMathOperator{\tr}{tr}
\DeclareMathOperator{\Tr}{Tr}
\DeclareMathOperator{\diag}{diag}
\def\chargec{\mathrm{C}}        
        
\newcommand{\RaiseBrace}[1]{\raise1.5pt\hbox{$\displaystyle#1$}}
\newcommand{\ChargeConjugate}[1]{{#1}^\chargec}
\newcommand{\irrep}[1]{\ensuremath{\bf #1}}
\newcommand{\EWrep}[2]{\ensuremath{\left(\irrep{#1},#2\right)}}
\newcommand{\hc}{\ensuremath{\text{h.c.}}}

\newcommand{\DeltaBarSF}{\ensuremath{\overline{\SuperField{\Delta}}}}

\allowdisplaybreaks[1]

\begin{document}
%\begin{fmffile}{rgeTripletFMF}
\received{06/19/2007}

\title{Renormalization Group Evolution in the type I + II seesaw model}
\author{Michael Andreas Schmidt}
\email{michael.schmidt@ph.tum.de}
\affiliation{MPI f\"{u}r Kernphysik, Saupfercheckweg 1, 69117 Heidelberg, Germany}

\preprint{arXiv: 0705.3841 [hep-ph]}

\begin{abstract}
\noindent
We carefully analyze the renormalization group equations in the type I +
II seesaw scenario in the extended standard model (SM) and minimal
supersymmetric standard model (MSSM). Furthermore, we present
analytic formulae of the mixing angles and phases and discuss the RG
effect on the different mixing parameters in the type II seesaw
scenario. The renormalization group equations of the angles have a contribution which is proportional to the mass squared
difference for a hierarchical spectrum. This is in contrast to the inverse
proportionality to the mass squared difference in the effective field
theory case.
\end{abstract}

\maketitle

\section{Introduction}
In the last years, the precision of the leptonic mixing parameters has been further increased~\cite{Schwetz:2006dh} due to a number of different experiments~\cite{Miuchi:2003cm,Araki:2004mb,Aliu:2004sq,Aharmim:2005gt,Ashie:2005ik,Rebel:2007ex}. 
The precision of measurements of leptonic mixing parameters will even increase in upcoming experiments. In the next-generation experiments, the mixing parameters will be measured on a 10 \% level~\cite{Huber:2004ug}. Thus we are entering an era of precision experiments in neutrino physics.

Contrarily, there have been a lot of attempts to explain the structure of the neutrino mass
matrix ({\it e.g.} ~\cite{Altarelli:2004za,Mohapatra:2005wg,Mohapatra:2006gs}). However, most models use heavy particles which generate light neutrino masses effectively after decoupling. Examples are right--handed neutrinos in the standard (type I) seesaw mechanism~\CiteSeeSaw\ 
\begin{equation}\label{eq:SeeSaw}
 m_\nu\,=\,- (m_\nu^\mathrm{Dirac})^T\,M^{-1}\,m_\nu^\mathrm{Dirac}
\end{equation}
or a Higgs triplet in the type II seesaw mechanism~\cite{Magg:1980ut,Lazarides:1980nt,Mohapatra:1980ia}
\begin{equation}\label{eq:SeeSawTriplet}
    m_\nu= v_\Delta Y_\Delta\; ,
  \end{equation}
where $v_\Delta$ is the vacuum expectation value (vev) of the Higgs triplet and $Y_\Delta$ is the Yukawa coupling matrix of the vertex $\ell\Delta\ell$.
Clearly, the seesaw operates at high
energy scales while its implications are measured by experiments at low scales.
Therefore, the neutrino masses given by Eqs.~(\ref{eq:SeeSaw},\ref{eq:SeeSawTriplet}) are  subject to
quantum corrections, i.e.\ they are modified by renormalization group (RG)
running.

The running of neutrino masses and leptonic mixing angles has 
been studied intensely in the literature. 
RG effects can be very large for a quasi-degenerate neutrino mass hierarchy and they can have interesting implications for model building. For instance, leptonic mixing angles can be magnified ~\cite{Balaji:2000au,Miura:2000bj,Antusch:2002fr,Mohapatra:2003tw,Hagedorn:2004ba}, the small mass splittings can be generated from exactly degenerate light neutrinos 
~\cite{Chankowski:2000fp,Chun:2001kh,Chen:2001gk,Joshipura:2002xa,Joshipura:2002gr,Singh:2004zu}
or bimaximal mixing at high energy can be made compatible with low-energy experiments ~\cite{Antusch:2002hy,Miura:2003if,Shindou:2004tv}. 
On the other hand, even rather small RG corrections are important in
view of the precision era we are entering. For example, RG effects induce deviations
from $\theta_{13}=0$ or maximal mixing $\theta_{23}=\pi/4$
~\cite{Antusch:2003kp,Mei:2004rn,Antusch:2004yx} also for a hierarchical
spectrum, as well as from other symmetries, like
quark-lepton-complementarity~\cite{Minakata:2004xt,Dighe:2006zk,Schmidt:2006rb,Dighe:2007ks},
tribimaximal
mixing~\cite{Plentinger:2005kx,Luo:2005fc,Mohapatra:2006pu,Hirsch:2006je,Dighe:2006sr,Dighe:2007ks}
and other special
configurations~\cite{Mei:2005gp,Xing:2006vk}. Threshold corrections can
yield large effects~\cite{Tanimoto:1995bf,Casas:1999tp,Casas:1999ac,King:2000hk,Antusch:2002rr,%
Antusch:2002hy,Antusch:2002fr,Miura:2003if,Shindou:2004tv,Mei:2004rn,Antusch:2005gp,Hirsch:2006je,Schmidt:2006rb}.

These studies have been done in the effective theory of Majorana neutrino masses~\cite{Chankowski:1993tx,Babu:1993qv,Casas:1999tg,Antusch:2003kp,Mei:2004rn,Antusch:2004yx,Luo:2006tb}. There are also studies in the standard seesaw case~\cite{Antusch:2005gp,Mei:2005qp} and in the case of Dirac neutrinos~\cite{Lindner:2005as,Xing:2006sp} which also show significant RG effects which can become comparable to the precision of experimental data. Therefore the RG effects have to be considered in model building in order to be able to compare predictions to experimental data.

In this paper we present the RG equations in the type II seesaw
scenario~\footnote{Recently Chankowski et al. showed that a $Y=0$ Higgs triplet does
  not decouple from the SM~\cite{Chankowski:2006hs} which imposes strong
  constraints on models involving a $Y=0$ Higgs triplet. However, this result
  can not immediately translated to a $Y=1$ Higgs triplet like in the type-II
  seesaw mechanism, because the non-vanishing hypercharge leads to a coupling of
  the Higgs triplet to the hypercharge boson, which in turn leads to different
  contributions to the electroweak precision observables, e.g. the mass of the
  $Z$ boson receives additional corrections which results in corrections to the
  $\rho=\frac{M_W^2}{M_Z^2\cos^2\theta_W}$ parameter.} in the standard
model (SM) ~\cite{MichaelDip} and minimal supersymmetric standard model (MSSM). 
We derive analytic formulae which allow to understand  the running of
the neutrino parameters above the threshold of the Higgs
triplet. Furthermore, we extend the software package
REAP/MixingParameterTools~\footnote{REAP/MixingParameterTools can be
  downloaded from \url{http://www.ph.tum.de/\~rge}.} by an Higgs triplet
for analyzing the RG evolution numerically.
A similar calculation has been done by Chao and Zhang on the
renormalization of the SM extended by a Higgs
triplet~\cite{Chao:2006ye}. The two calculations differ in several
terms~\footnote{The calculation of the necessary diagrams in the
  extended SM can be downloaded from
  \url{http://www.mpi-hd.mpg.de/\~mschmidt/rgeTriplet}. Chao and Zhang
  checked the relevant parts of their calculation and obtained the same
  results as we do.}. 

The paper is organized as follows: In Sec.~\ref{sec:Lagrangian}, we present the Lagrangian of the type II seesaw model and give the tree--level matching conditions. Furthermore in Sec.~\ref{sec:counterterms} we show all new wave function renormalization factors and counterterms and point out the differences to the work of Chao and Zhang~\cite{Chao:2006ye}. The RG equations are shown in Sec.~\ref{sec:RGE}. In Sec.~\ref{sec:MSSM}, the additional terms in the superpotential and the RG equations in the MSSM are presented\footnote{The case without right--handed neutrinos has been studied in~\cite{Rossi:2002zb}.}. Sec.~\ref{sec:MixingParameters} is dedicated to the analytic understanding of RG effects in the type II seesaw case (only a Higgs triplet) and Sec.~\ref{sec:fullMP} gives a glimpse on the full type I + II seesaw case. Finally, Sec.~\ref{sec:Conclusions} contains our conclusions.

\section{Type II Seesaw Lagrangian\label{sec:Lagrangian}}

In the following, we consider the SM extended by right-handed neutrinos $\nu_R\sim\EWrep{1}{0}_{\SU(2)\times\U(1)}$ and a charged Higgs triplet $\Delta\sim\EWrep{3}{1}_{\SU(2)\times\U(1)}$,
\begin{equation}
\Delta=\frac{\sigma^i}{\sqrt{2}}\Delta_i=\left(\begin{array}{cc}
    \Delta^+/\sqrt{2} & \Delta^{++}\\
    \Delta^0 & -\Delta^+/\sqrt{2}\\
  \end{array}\right)\; .
\end{equation}
The Lagrangian is given by
\begin{equation}
\mathscr{L} = \mathscr{L}_\text{SM} + \mathscr{L}_{\nu_R} + \mathscr{L}_\Delta \; ,
\end{equation}
where the individual parts are defined by
\begin{subequations}
  \begin{align}
  \mathscr{L}_{\nu_R}=&\overline{\nu_R} i \fmslash{\partial}\nu_R-\left(Y_\nu\right)_{ij}\overline{\nu_R}^i \tilde\Higgs^\dagger \ell_L^j%\ChargeConjugate{\Higgs}
  -\frac{1}{2}M_{ij}\overline{\ChargeConjugate{\nu_R}}^i \nu_R^j+\hc\\
  \mathscr{L}_\Delta=&\tr\left[\left(D_\mu \Delta\right)^\dagger D^\mu\Delta\right] - \mathscr{V}(\Delta, \Higgs)\nonumber\\&-\frac{1}{\sqrt{2}}\left(Y_\Delta\right)_{fg}\ell_L^{Tf} \mathrm{C} i\sigma_2\Delta \ell_L^g +\hc\\
\mathscr{V}(\Delta, \Higgs) = & M_\Delta^2\tr\left(\Delta^\dagger \Delta\right) 
+ \frac{\Lambda_1}{2}\left(\tr\Delta^\dagger\Delta\right)^2 \nonumber\\
&+ \frac{\Lambda_2}{2}\left[\left(\tr\Delta^\dagger\Delta\right)^2-\tr\left(\Delta^\dagger\Delta\Delta^\dagger\Delta\right)\right] \nonumber\\
&+ \Lambda_4\Higgs^\dagger\Higgs\tr\left(\Delta^\dagger\Delta\right) + \Lambda_5\Higgs^\dagger\left[\Delta^\dagger,\Delta\right]\Higgs \nonumber\\
&+ \left[ \frac{\Lambda_6}{\sqrt{2}} \Higgs^T i\sigma_2 \Delta^\dagger \Higgs +\hc \right]\label{eq:HHDelta}
\end{align}
\end{subequations}
The covariant derivative of the Higgs triplet is given by~\footnote{We use GUT charge normalization: $\frac{3}{5} \left(g_1^\mathrm{GUT}\right)^2 = \left(g_1^\mathrm{SM}\right)^2$}
\begin{equation}
D_\mu \Delta = \partial_\mu \Delta +i \sqrt{\frac{3}{5}} g_1 B_\mu \Delta + i g_2 \left[ W_\mu,\Delta\right]\;,
\end{equation}
$\tilde\phi=i\sigma_2 \phi^*$, 
and $\mathrm{C}$ is the charge conjugation matrix with respect to the Lorentz group.
The counterterm parts of the Lagrangian which are needed in the paper are given in App.~\ref{app:counterterm}.
In addition, we consider an effective dimension 5 (D5) operator which
generates neutrino masses because it appears as soon as the Higgs triplet or a right--handed neutrino decouples:
\begin{equation}
\mathscr{L}_\kappa = -\frac{1}{4}\kappa_{fg} (\overline{\ell_L^{C}}^fi\sigma_2\Higgs) (\ell_L^g i\sigma_2\Higgs)\; .
\end{equation}
The most general neutrino mass matrix is given by the following formula
\begin{equation}
    m_\nu=-\frac{v^2}{4}\left(\kappa+2 Y_\nu^T M^{-1} Y_\nu-2\frac{Y_\Delta\Lambda_6}{M_\Delta^2}\right)\; ,
\end{equation}
where $\kappa$ includes additional contributions to the dimension 5 operator, like from gravitational effects~\cite{Wetterich:1981bx}. Thus the $\beta$-function of the neutrino mass is given by the sum of the $\beta$-functions for the
contribution from the right-handed neutrinos and the contribution from the Higgs triplet.

The right--handed neutrinos and the Higgs triplet decouple step by step
at their respective mass scale and the effective theories have to be
matched against each other. The decoupling of the right--handed
neutrinos only contributes to the effective D5 operator. The decoupling of the Higgs triplet also gives a contribution to the SM model Higgs self--coupling because there is a coupling between the SM Higgs doublet and the Higgs triplet given in Eq.\eqref{eq:HHDelta}.
Hence, the matching conditions of the right-handed neutrinos are
\begin{equation}
  \kappa^\mathrm{EFT}=\kappa+2Y_\nu^T M^{-1} Y_\nu
\end{equation}
and the decoupling of the Higgs triplet leads to
\begin{subequations}
\begin{align}
\kappa^\mathrm{EFT} &= \kappa - 2 \frac{Y_\Delta \Lambda_6}{M_\Delta^2}\\
\lambda^\mathrm{EFT} &= \lambda + 2\frac{|\Lambda_6|^2}{M_\Delta^2}\; .
\end{align}
\end{subequations}
In the following two sections, we present the calculation of the renormalization group equations. The calculation has been carefully double checked and compared to the results of Chao and Zhang~\cite{Chao:2006ye}.
\section{Wave function renormalization and
  Counterterms\label{sec:counterterms}}

\begin{figure}
% \subfigure{
% $\fmfframe(10,4)(4,4){    
% \begin{fmfgraph*}(24,18)
% \fmfleft{l1,l2}
% \fmfright{r}
% \fmf{scalar}{l1,v1}
% \fmf{scalar}{l2,v1}
% \fmf{scalar}{v3,r}
% \fmf{scalar,label=$\Higgs$,tension=.4,left}{v1,v3}
% \fmf{scalar,label=$\Higgs$,tension=.4,right}{v1,v3}
% \fmffreeze
% \fmfdot{v1,v3}
% \fmflabel{$\Higgs$}{l2}
% \fmflabel{$\Higgs$}{l1}
% \fmflabel{$\Delta$}{r}
% \end{fmfgraph*}}$
% }\quad
% \subfigure{
% $\fmfframe(10,4)(4,4){    
% \begin{fmfgraph*}(24,16)
% \fmfleft{l1,l2}
% \fmfright{r}
% \fmf{scalar}{l1,v1}
% \fmf{scalar}{l2,v3}
% \fmf{scalar}{v3,r}
% \fmf{scalar,label=$\Delta$,tension=.2,left}{v1,v3}
% \fmf{scalar,label=$\Higgs$,tension=.2,left}{v3,v1}
% \fmffreeze
% \fmfdot{v1,v3}
% \fmflabel{$\Higgs$}{l2}
% \fmflabel{$\Higgs$}{l1}
% \fmflabel{$\Delta$}{r}
% \end{fmfgraph*}}$
%}
\subfigure{\includegraphics{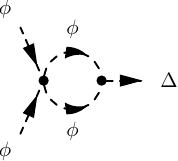}}\quad
\subfigure{\includegraphics{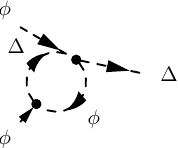}}
\caption{\label{fig:feynman}Feynman diagrams which are not considered in \cite{Chao:2006ye}.}
\end{figure}

In order to obtain the wave function renormalization factor, all
self--energy diagrams of the involved particles have to be calculated to
one loop order. We use dimensional regularization together with the
$\overline{MS}$ scheme, because gauge invariance is generically preserved in this scheme.
The differences to the formulae in ~\cite{Chao:2006ye} are underlined. Singly underlined terms differ in the prefactor and wavy underlined terms are not present in ~\cite{Chao:2006ye}. Thus they probably stem from diagrams not taken into account in ~\cite{Chao:2006ye}. For example, the wavy underlined terms in the counterterm of $\Lambda_6$ are due to the Feynman diagrams shown in Fig.\ref{fig:feynman}.
% \end{equation}
%
Our Feynman rules and calculation can be downloaded from \url{http://www.mpi-hd.mpg.de/\~mschmidt/rgeTriplet/}. 
As we use a different definition for the couplings in the Lagrangian as in ~\cite{Chao:2006ye}, the translation is presented in Tab.~\ref{tab:translation}.
\begin{table*}
\begin{tabular}{|l||l|l|l|l|l|l|l|l|l|}\hline
our work & $Y_\Delta$ & $m$ & $M_\Delta$ & $\lambda$ & $\Lambda_1$ & $\Lambda_2$ & $\Lambda_4$ & $\Lambda_5$ & $\Lambda_6$\\\hline
Chao and Zhang~\cite{Chao:2006ye} & $\frac{1}{\sqrt{2}}Y_\xi$ & $m_\phi$ & $M_\xi$ & $\lambda$ & $\frac{1}{2}\lambda_\xi$ & $\lambda_C$ & $\lambda_\phi$ & $-\frac{1}{2}\lambda_T$ & $-\frac{1}{\sqrt{2}}M_\xi \lambda_H$\\\hline
\end{tabular}
\caption{Translation table for all relevant parameters}
\label{tab:translation}
\end{table*}
The wave function renormalization factors are given by
\begin{subequations}
\begin{align}
  \delta Z_\Delta=&
  \frac{1}{16\pi^2\epsilon}\left[\frac{6}{5}\left(3-\xi_1\right)g_1^2+4\left(3-\xi_2\right)g_2^2-2\tr\left(Y_\Delta^\dagger Y_\Delta\right)\right]\\
  \delta Z_\Higgs=&-\frac{1}{16\pi^2\epsilon}\left[2T-\frac{3}{10}\left(3-\xi_1\right)g_1^2-\frac{3}{2}\left(3-\xi_2\right)g_2^2\right]\\
  \delta Z_{\ell_L}=&-\frac{1}{16\pi^2\epsilon}\left[Y_\nu^\dagger Y_\nu+Y_e^\dagger
  Y_e+3Y_\Delta^\dagger Y_\Delta\right]\nonumber\\
&-\frac{1}{16\pi^2\epsilon}\left[\frac{3}{10}\xi_1g_1^2+\frac{3}{2}\xi_2g_2^2\right]\label{eq:ellL}\; ,
 \end{align}
\end{subequations}
where we have defined
\begin{equation}
T= \tr\left(Y_\nu^\dagger Y_\nu+Y_e^\dagger Y_e+3 Y_u^\dagger Y_u+3Y_d^\dagger Y_d\right)\; .
\end{equation}
The Yukawa coupling vertex $\ell\Delta\ell$ can be renormalized multiplicatively
\begin{equation}
  \delta Z_{Y_\Delta}=\frac{1}{32\pi^2\epsilon}\left[\frac{9}{5}\left(1-\xi_1\right)g_1^2+\left(3-7\xi_2\right)g_2^2\right]
\end{equation}
the parameters in the Higgs potential, however, have to be renormalized additively
\begin{widetext}
\begin{subequations}
\begin{align}
  \delta m^2=&\frac{1}{16\pi^2\epsilon}\left[\left(3\lambda
  -\frac{3}{10}\xi_1g_1^2-\frac{3}{2}\xi_2g_2^2\right)m^2-4\tr\left(Y_\nu^\dagger
  M_\Delta^2 Y_\nu\right)+6\Lambda_4 M_\Delta^2+6|\Lambda_6|^2\right]\\
\delta M_\Delta^2 =& \frac{1}{16\pi^2\epsilon}\left[\left(8\Lambda_1+2\Lambda_2- \frac{6}{5}\xi_1
 g_1^2-4\xi_2 g_2^2\right)M_\Delta^2 +4\Lambda_4 m^2 +2|\Lambda_6|^2\right]\\
    \delta \lambda =&\frac{1}{16\pi^2\epsilon}\bigg[6\lambda^2-\frac{1}{2}\lambda\left(\frac{3}{5}g_1^2+3g_2^2\right)+3 g_2^4+\frac{3}{2}\left(\frac{3}{5}g_1^2+g_2^2\right)^2\nonumber\\
&-8\tr\left[Y_e^\dagger Y_eY_e^\dagger Y_e +Y_\nu^\dagger Y_\nu Y_\nu^\dagger Y_\nu +3 Y_u^\dagger Y_uY_u^\dagger Y_u + 3 Y_d^\dagger Y_dY_d^\dagger Y_d\right] + 12 \Lambda_4^2+ 8\Lambda_5^2\bigg]\\
  \delta\Lambda_1=&\frac{1}{16\pi^2\epsilon}\left[\left(-\frac{12}{5}\xi_1g_1^2-8g_2^2\xi_2\right)\Lambda_1 +\frac{9}{25} 12 g_1^4 \uline{+18g_2^4}+\frac{72}{5}g_1^2g_2^2\uline{+14\Lambda_1^2}+2\Lambda_2^2+4\Lambda_1\Lambda_2+4\left(\Lambda_4^2+\Lambda_5^2\right)\right.\nonumber\\
&\left.\uline{-8\tr\left(Y_\Delta^\dagger Y_\Delta Y_\Delta^\dagger Y_\Delta\right)}\right]\\
  \delta\Lambda_2=&\frac{1}{16\pi^2\epsilon}\left[\left(-\frac{12}{5}\xi_1g_1^2-8g_2^2\xi_2\right)\Lambda_2\uwave{+12g_2^4}-\frac{144}{5}g_1^2g_2^2+3\Lambda_2^2+12\Lambda_1\Lambda_2-8\Lambda_5^2\uline{+8\tr\left(Y_\Delta^\dagger Y_\Delta Y_\Delta^\dagger Y_\Delta\right)}\right]\\
       \delta\Lambda_4=&\frac{1}{16\pi^2\epsilon}\Bigg[\left(-\frac{3}{2}\xi_1g_1^2-\frac{11}{2}\xi_2g_2^2\right)\Lambda_4+\frac{27}{25}g_1^4\uline{+6g_2^4}+\left(8\Lambda_1+2\Lambda_2+3\lambda+4\Lambda_4\right)\Lambda_4+8\Lambda_5^2
       \\\nonumber&
       -4\tr\left(Y_\Delta^\dagger Y_\Delta Y_\nu^\dagger Y_\nu\right)
       -4\tr\left(Y_\Delta^\dagger Y_\Delta Y_e^\dagger Y_e\right)
       \Bigg]\\
       \delta\Lambda_5=&\frac{1}{16\pi^2\epsilon}\Bigg[\left(-\frac{3}{2}\xi_1g_1^2-\frac{11}{2}\xi_2g_2^2\right)\Lambda_5-\frac{18}{5}g_1^2g_2^2+\left(2\Lambda_1-2\Lambda_2+\lambda+8\Lambda_4\right)\Lambda_5
       \\\nonumber&
       +4\tr\left(Y_\Delta^\dagger Y_\Delta Y_\nu^\dagger Y_\nu\right)
       -4\tr\left(Y_\Delta^\dagger Y_\Delta Y_e^\dagger Y_e\right)
       \Bigg]\\
	\delta\Lambda_6=&-\frac{1}{16\pi^2\epsilon}\left[-4\tr\left(Y_\Delta^\dagger Y_\nu^T M Y_\nu \right)+\left(\frac{9}{10}g_1^2\xi_1+\frac{7}{2}g_2^2\xi_2\uwave{-\lambda-4\Lambda_4+8\Lambda_5}\right)\Lambda_6\right]\; .
\end{align}
\end{subequations}
\end{widetext}

\section{Renormalization Group Equations\label{sec:RGE}}
Using the vertex corrections and the wave function renormalization factors, we can deduce the $\beta$--functions via the formula given in ~\cite{Antusch:2001ck}. 
As the wave function renormalization constant for the left-handed lepton doublets
has an additional term with respect to the SM extended by right-handed neutrinos, all vertices receive an additional contribution per
left-handed lepton attached to the vertex which is given by
\begin{equation}
  \frac{3}{2} \frac{1}{16\pi^2} Y_\Delta^\dagger Y_\Delta
\end{equation}
multiplied by the matrix characterizing the vertex from the left or from the
right, respectively.
In particular, the $\beta$-functions of the lepton Yukawa
couplings~\cite{JoernDip,StefanPhD, MichaelPhD} and the Yukawa coupling $Y_\Delta$ become

\begin{widetext}
\begin{subequations}
\begin{align}
     16\pi^2\Dot Y_\nu=&Y_\nu\left[\frac{3}{2}Y_\nu^\dagger Y_\nu-\frac{3}{2}Y_e^\dagger Y_e+\frac{3}{2}Y_\Delta^\dagger Y_\Delta\right] + Y_\nu\left[T-\frac{9}{20}g_1^2-\frac{9}{4}g_2^2\right]\\
16\pi^2\Dot Y_e=&Y_e\left[\frac{3}{2}Y_e^\dagger Y_e-\frac{3}{2}Y_\nu^\dagger Y_\nu+\frac{3}{2}Y_\Delta^\dagger Y_\Delta\right] +Y_e\left[T-\frac{9}{4}g_1^2-\frac{9}{4}g_2^2\right]\\
  16\pi^2\Dot Y_\Delta=&\left[\frac{1}{2}Y_\nu^\dagger Y_\nu+\frac{1}{2}Y_e^\dagger Y_e+\frac{3}{2}Y_\Delta^\dagger Y_\Delta\right]^T Y_\Delta +Y_\Delta\left[\frac{1}{2}Y_\nu^\dagger Y_\nu+\frac{1}{2}Y_e^\dagger  Y_e+\frac{3}{2}Y_\Delta^\dagger Y_\Delta\right] +\left[-\frac{3}{2}\left(\frac{3}{5}g_1^2+3g_2^2\right)+\tr\left(Y_\Delta^\dagger
  Y_\Delta\right)\right]Y_\Delta\; .
\end{align}
\end{subequations}
The renormalization of $\Lambda_6$ is described by
  \begin{align}
  16\pi^2\Dot \Lambda_6=&\Big[\uwave{\lambda+4\Lambda_4-8\Lambda_5}-\frac{27}{10}g_1^2-\frac{21}{2}g_2^2
	  +2 T +  \tr\left(Y_\Delta^\dagger Y_\Delta\right)\Big]\Lambda_6 +4\tr\left( Y_\Delta^\dagger Y_\nu^T M Y_\nu\right)
\end{align}
and the anomalous dimensions of the Higgs triplet mass term is given by
\begin{align}
  16\pi^2\gamma_{M_\Delta}=&\frac{9}{5}g_1^2+6g_2^2-4\Lambda_1-\Lambda_2-\tr\left(Y_\Delta^\dagger
	Y_\Delta\right)+\left(-2\Lambda_4 m^2-|\Lambda_6|^2\right)M_\Delta^{-2}\; .
\end{align}
The $\beta$-function of the effective neutrino mass operator $\kappa$ function changes to
\begin{align}
  16\pi^2\Dot\kappa=&\left[\frac{1}{2}Y_\nu^\dagger Y_\nu -\frac{3}{2}Y_e^\dagger Y_e+\frac{3}{2}Y_\Delta^\dagger Y_\Delta\right]^T\kappa
+\kappa\left[\frac{1}{2}Y_\nu^\dagger Y_\nu -\frac{3}{2}Y_e^\dagger Y_e+\frac{3}{2}Y_\Delta^\dagger Y_\Delta\right]
+\left[2 T -3g_2^2+\lambda\right]\kappa\; .
\end{align}
\end{widetext}

The RG equation for the type I contribution to the neutrino mass is only
changed by the additional term to the $\beta$-function of the neutrino
Yukawa couplings due to the Higgs triplet.
The remaining RG equations are presented in App.~\ref{app:RGE}. They either do not receive additional contributions or do not directly influence the neutrino mass matrix.
The main difference in the RG equations compared to the results
in~\cite{Chao:2006ye} arise from contributions of the additional diagrams contributing to $\Lambda_6$. As it can been seen later, they have an impact on the evolution of neutrino masses, but the evolution of mixing angles and phases remains unchanged. 
In summary, the running of the effective neutrino mass matrix $m_\nu$  above and
between the seesaw scales  is given by the running of the three different contributions to the neutrino mass matrix,
\begin{align}\label{eq:EffNuMass}
m_\nu^{(1)} &= -\frac{v^2}{4} \kappa, \nonumber\\
m_\nu^{(2)}&= -\frac{v^2}{2} Y_\nu^T M^{-1} Y_\nu, \nonumber\\ 
m_\nu^{(3)}&=\frac{v^2}{2} \Lambda_6 M_\Delta^{-2} Y_\Delta\; .
\end{align}
The 1-loop $\beta$-functions for $m_\nu$  in the various
effective theories can be summarized as
\begin{align}
16 \pi^2 \, 
\frac{\D m_\nu^{(i)}}{\D t} = &\left[C_e Y_e^\dagger Y_e + C_\nu Y^\dagger_\nu   
   Y_\nu + C_\Delta Y_\Delta^\dagger Y_\Delta\right]^T \:m_\nu^{(i)} \nonumber\\ 
&+ m_\nu^{(i)} \, \left[ C_e Y_e^\dagger Y_e + C_\nu Y^\dagger_\nu Y_\nu + C_\Delta Y_\Delta^\dagger Y_\Delta \right]\nonumber\\
 &+ \alpha\, m_\nu^{(i)} \; ,\label{eq:BetaEffNuMass}
\end{align}
where $m_\nu^{(i)}$ stands for any of the three contributions to the neutrino mass matrix, 
respectively. The coefficients $C_{e,\nu,\Delta}$ and $\alpha$ are listed in  
Tab.~\ref{tab:BetaEffNuMass}. In the type-I+II seesaw scenario, large RG effects can be expected before the Higgs triplet is integrated out due to the different coefficients ($C_e$, $C_\nu$, $C_\Delta$) in analogy to the standard seesaw scenario where large RG corrections between the thresholds are induced by additional flavor--diagonal vertex corrections to the D5 operator.

\begin{table*}
\begin{tabular}{|l|l||c|c|c|p{9cm}|}
\hline
model & $m_\nu^{(i)}$ $\vphantom{\frac{1}{2}}$ &
$\!C_e\!$ & $\!C_\nu\!$ & $\!C_\Delta\!$ & flavor-trivial term $\alpha$\\
\hline 
SM & $\kappa$ \(\vphantom{\sqrt{\big|}^C}\)&   
$\!-\tfrac{3}{2}\!$ & $\tfrac{1}{2}$ & $\!\tfrac{3}{2}$ &
\begin{minipage}{9cm}
$2 T -3 g_2^2+\lambda$
\end{minipage}
\\
SM & $2\, Y_\nu^T M^{-1}
Y_\nu \!\!$ 
$\vphantom{\frac{1}{2}}$\(\vphantom{\sqrt{\big|}^C}\)&
$\!-\tfrac{3}{2}\!$ & $\tfrac{1}{2}$ & $\!\tfrac{3}{2}$ &
\begin{minipage}{9cm}
$2 T 
  -\tfrac{9}{10} g_1^2 - \tfrac{9}{2} g_2^2$
\end{minipage}
\\
SM & $-2\, \Lambda_6 M_\Delta^{-2}Y_\Delta$ 
$\vphantom{\frac{1}{2}}$\(\vphantom{\sqrt{\big|}^C}\)&
$\!\tfrac{1}{2}\!$ & $\tfrac{1}{2}$ & $\tfrac{3}{2}$ &
\begin{minipage}{9cm}
$
2T -3g_2^2+\lambda-8\Lambda_1-2\Lambda_2 +4\Lambda_4-8\Lambda_5 
-\left(4 \Lambda_4 m^2 +2 |\Lambda_6|^2\right) M_\Delta^{-2}+4 \tr\left(Y_\Delta^\dagger Y_\nu^T M Y_\nu\right)\Lambda_6^{-1}
$
\end{minipage}\\
\hline 
%%%%%%%%%%%%%%%%%%%%%%%%%%%%%%%%%%%%%%%%%%%5
MSSM &$\kappa$ \(\vphantom{\sqrt{\big|}^C}\)&   
$1$ & $1$ & $3$ &
\begin{minipage}{9cm}
$2 \tr\left(Y_\nu^\dagger Y_\nu+3 Y_u^\dagger Y_u\right) +6\,|\Lambda_u|^2 -2 \left(\frac{3}{5} g_1^2 + 3 g_2^2 \right)$
\end{minipage}
\\
MSSM &$2\, Y_\nu^T M^{-1}
Y_\nu \!\!$ 
$\vphantom{\frac{1}{2}}$\(\vphantom{\sqrt{\big|}^C}\)&
$1$ & $1$ & $3$ &
\begin{minipage}{9cm}
$2 \tr\left(Y_\nu^\dagger Y_\nu+3 Y_u^\dagger Y_u\right) +6\,|\Lambda_u|^2 -2 \left(\frac{3}{5} g_1^2 + 3 g_2^2 \right)$
\end{minipage}
\\
MSSM &$-2\, \Lambda_u M_\Delta^{-1}Y_\Delta$ 
$\vphantom{\frac{1}{2}}$\(\vphantom{\sqrt{\big|}^C}\)&
$1$ & $1$ & $3$ &
\begin{minipage}{9cm}
$2 \tr\left(Y_\nu^\dagger Y_\nu+3 Y_u^\dagger Y_u\right)+6\,|\Lambda_u|^2 -2 \left(\frac{3}{5} g_1^2 + 3 g_2^2 \right)$
\end{minipage}
\\
\hline
\end{tabular}
\caption{Coefficients of the $\beta$-functions of Eq.~\eqref{eq:BetaEffNuMass}, which govern the running of the effective neutrino mass matrix in minimal type II seesaw models. In the MSSM, the coefficients coincide due to the non-renormalization theorem~\cite{Wess:1973kz,Iliopoulos:1974zv} in supersymmetric theories.}
\label{tab:BetaEffNuMass}
\end{table*}

\section{Higgs triplet in the MSSM\label{sec:MSSM}}
\begin{figure}[htb]
\subfigure{\includegraphics[width=3.5cm]{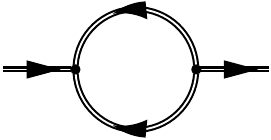}}\quad\quad
\subfigure{\includegraphics[width=3.5cm]{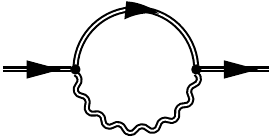}}
\caption{Supergraphs which contribute to the wave function renormalization.}
\label{fig:supergraphs}
\end{figure}

In the MSSM, in addition to the Higgs triplet $\SuperField{\Delta}\sim\EWrep{3}{1}$, a second Higgs triplet $\DeltaBarSF\sim\EWrep{3}{-1}$ with opposite hypercharge $Y$ is needed to generate a D5 mass term for neutrinos. Furthermore, $\DeltaBarSF$ ensures that the model is anomaly-free. Note, however, that only one Higgs triplet couples to the left--handed leptons. The additional terms in the superpotential are given by
\begin{equation}
\begin{split}
W_\Delta = &M_\Delta \Tr(\DeltaBarSF\SuperField{\Delta})  + \frac{\left(Y_\Delta\right)_{fg}}{\sqrt{2}} \SuperField{l}^{fT} i\sigma_2 \SuperField{\Delta} \SuperField{l}^{g}\\
&+ \frac{\Lambda_u}{\sqrt{2}} {\SuperField{h}^{(2)}}^T i\sigma_2\DeltaBarSF \SuperField{h}^{(2)}+ \frac{\Lambda_d}{\sqrt{2}}{\SuperField{h}^{(1)}}^T i\sigma_2\SuperField{\Delta}\SuperField{h}^{(1)}\; ,
\end{split}
\end{equation}
where $\SuperField{l}$ denotes the left-handed doublet and $\SuperField{h}^{(i)}$ denotes the Higgs doublets. We use the same notation as in ~\cite{Antusch:2002ek}.
The decoupling of the Higgs triplet generates an effective dimension 4
term $\kappa^\mathrm{EFT}$ in the superpotential, whereas the tree-level matching condition reads
\begin{equation}
\kappa^\mathrm{EFT} = \kappa - 2 \frac{Y_\Delta \Lambda_u}{M_\Delta}\; .
\end{equation}
The RG equations~\footnote{The
  terms coming from the Higgs triplet have been obtained earlier by Rossi~\cite{Rossi:2002zb}.} can be obtained easily by
using the supergraph technique as it is described in
~\cite{Antusch:2002ek}. There are only two different types of
supergraphs contributing to the wave function renormalization which are
shown in Fig.~\ref{fig:supergraphs}.
Here, we only show the RG equations which are relevant for the RG evolution of the neutrino mass matrix
\begin{widetext}
\begin{align}
16\pi^2\Dot Y_\Delta = &Y_\Delta \left[Y_e^\dagger Y_e + Y_\nu^\dagger Y_\nu +3Y_\Delta^\dagger Y_\Delta \right] +\left[Y_e^\dagger Y_e + Y_\nu^\dagger Y_\nu +3Y_\Delta^\dagger Y_\Delta \right]^T Y_\Delta + Y_\Delta\left[|\Lambda_d|^2+\tr(Y_\Delta^\dagger Y_\Delta)-\frac{9}{5}g_1^2-7g_2^2\right]\\
16\pi^2\Dot\Lambda_u=&\Lambda_u\left[2\tr(3Y_u^\dagger Y_u+Y_\nu^\dagger Y_\nu)+7\,|\Lambda_u|^2-\frac{9}{5}g_1^2-7g_2^2\right]\\
16\pi^2\Dot M_\Delta= &M_\Delta\left[\tr(Y_\Delta^\dagger Y_\Delta)+|\Lambda_u|^2+|\Lambda_d|^2-4\left(\frac{3}{5}g_1^2+2g_2^2\right)\right]\; .
\end{align}
As it can be seen in the $\beta$-function of $Y_\Delta$ for example, the sign of $Y_\Delta^\dagger Y_\Delta$ in $P$ equals the one in the SM which leads to the same sign in the RG equations of the angles in the limit of a strong hierarchy (See Sec.~\ref{sec:MixingParameters}).
The RG equation of the effective D5 operator $\kappa$ is
\begin{align}
16\pi^2\Dot \kappa = &\kappa \left[ Y_e^\dagger Y_e + Y_\nu^\dagger Y_\nu+3Y_\Delta^\dagger Y_\Delta\right] + \left[ Y_e^\dagger Y_e + Y_\nu^\dagger Y_\nu+3 Y_\Delta^\dagger Y_\Delta\right]^T \kappa + \kappa\left[2\tr(Y_\nu^\dagger Y_\nu +3 Y_u^\dagger Y_u)+6|\Lambda_u|^2-\frac{6}{5}g_1^2-6g_2^2\right]\; .
\end{align}
There is also an additional contribution to the RG equations which are relevant for the type I contribution to neutrino masses
\begin{align}
16\pi^2\Dot Y_\nu = &Y_\nu \left[ Y_e^\dagger Y_e + 3 Y_\nu^\dagger Y_\nu+3\, Y_\Delta^\dagger Y_\Delta\right] + Y_\nu\left[\tr(3 Y_u^\dagger Y_u + Y_\nu^\dagger Y_\nu)+3\,|\Lambda_u|^2-\left(\frac{3}{5}g_1^2+3g_2^2\right)\right]\\
16\pi^2\Dot M = &M \left(2 Y_\nu Y_\nu^\dagger \right)^T + \left(2 Y_\nu Y_\nu^\dagger \right) M\; .
\end{align}
\end{widetext}
All remaining RG equations are presented in App.~\ref{app:MSSM}.

The coefficients of the RG equations of the different contributions to
the neutrino mass matrix are summarized in
Tab.~\ref{tab:BetaEffNuMass}. Note, that the coefficients $C_e$,
$C_\nu$, $C_\Delta$ and $\alpha$ are the same for all three mass
contributions $m_\nu^{(i)}$ in the MSSM due to the non-renormalization theorem~\cite{Wess:1973kz,Iliopoulos:1974zv}.

\section{RG equations of mixing parameters in the type II seesaw case\label{sec:MixingParameters}}

In order to understand the RG evolution of neutrino masses and leptonic mixing parameters in the presence of a Higgs triplet, we consider a type II model, where the neutrino mass is generated by a Higgs triplet only. The evolution of the mixing parameters in standard parameterization can be described by the formulae of ~\cite{Antusch:2005gp} with suitable replacements for $P$, $F$, $\alpha$ and $\alpha_e$:
\begin{subequations}
\begin{align}
\Dot m_\nu = &P^T m_\nu + m_\nu P + \alpha\, m_\nu\\
\frac{\dd}{\dd t} Y_e^\dagger Y_e = & F^\dagger Y_e^\dagger Y_e + Y_e^\dagger Y_e F + \alpha_e Y_e^\dagger Y_e
\end{align}
\end{subequations} 
Here, we can express $P$ and $F$ in terms of physical parameters.
\begin{align}
P=&C_e\diag(y_e^2,y_\mu^2,y_\tau^2)+C_\Delta U^* \diag(y_1^2,y_2^2,y_3^2)U^T\label{eq:P}\\
F=&D_e\diag(y_e^2,y_\mu^2,y_\tau^2)+D_\Delta U^* \diag(y_1^2,y_2^2,y_3^2)U^T\label{eq:F}\; ,
\end{align}
where $U$ is the MNS matrix and $y_i=\frac{m_i}{v_\Delta}$.
We use the so-called standard-parameterization~\cite{Yao:2006px}
\begin{equation}
\begin{split}
 U  = &\diag(e^{\I\delta_{e}},e^{\I\delta_{\mu}},e^{\I\delta_{\tau}}) V\diag(e^{-\I\varphi_1/2},e^{-\I\varphi_2/2},1)\\
 V=  &R_{23}(\theta_{23}) \Gamma_\delta^\dagger R_{13}(\theta_{13})\Gamma_\delta R_{12}(\theta_{12}) \; ,
\end{split}
\end{equation}
where $R_{ij}$ is the matrix of rotation in the $i-j$ - plane and
$\Gamma_\delta=\diag(e^{\I\delta/2},1,e^{-\I\delta/2})$. Note, that the Majorana
phases drop out of the definition of $P$ and $F$ in flavor basis. 
We derive the RG equations by using the technique described in the appendix
of~\cite{Antusch:2005gp} which is based on earlier
works~\cite{Babu:1987im,Grzadkowski:1987tf,Casas:1999tg}. In the numerical
examples which are shown in the figures, we do not include any finite threshold
corrections, since we are considering 1 loop running and the finite threshold
corrections are assumed to be of the order of 2 loop RG running. Therefore, the Higgs triplet is
decoupled when its running mass equals the renormalization scale
\begin{equation}
\mu_\mathrm{dec}=M_\Delta(\mu_\mathrm{dec})\; .
\end{equation}
In all examples, we set $M_\Delta(\Lambda_\mathrm{GUT})=10^{10}\GeV$.
As we are only interested in showing the generic features of the RG
evolution, we choose the Higgs self-couplings to be
$\Lambda_{1,2,4,5}=0.5$ for simplicity, since they only indirectly
influence the RG evolution of the angles and the flavor-dependent part
of the RG equations of the masses. In a realistic model, the parameters
$\Lambda_i$ have to satisfy certain relations to produce the desired
vevs. 

In the following, we present all formulae in the approximation $y_e\ll y_\mu \ll y_\tau$ and $\theta_{13}\ll1$. The exact formulae can be downloaded from \url{http://www.mpi-hd.mpg.de/\~mschmidt/rgeTriplet}.

\subsection{Running of the masses}
The main contributions to the RG equations of the masses
\begin{subequations}
\begin{align}
16\pi^2 \frac{\Dot m_1}{m_1} = & \re \alpha + 2 C_\Delta \frac{m_1^2}{v_\Delta^2} + 2 C_e y_\tau^2 \sin^2\theta_{12}\sin^2\theta_{23}\nonumber\\&  +\Ord{\theta_{13}}\\
16\pi^2 \frac{\Dot m_2}{m_2} = & \re \alpha + 2 C_\Delta \frac{m_2^2}{v_\Delta^2} + 2 C_e y_\tau^2 \cos^2\theta_{12}\sin^2\theta_{23}\nonumber\\& +\Ord{\theta_{13}}\\
16\pi^2 \frac{\Dot m_3}{m_3} = & \re \alpha + 2 C_\Delta \frac{m_3^2}{v_\Delta^2} + 2 C_e y_\tau^2 \cos^2\theta_{23}  +\Ord{\theta_{13}}\label{eq:mDrei}
\end{align}
\end{subequations}
are the flavor--independent term $\re\alpha$ and the flavor--dependent term $2\, C_\Delta \frac{m_i^2}{v_\Delta^2}$. 
As the smallness of neutrino masses is usually explained by a small vev of the Higgs triplet $v_\Delta$, the singular values $y_i=\frac{m_i}{v_\Delta}$ of the Yukawa coupling $Y_\Delta$ can be of $\Ord{1}$. This in turn leads to sizable RG effects.
\begin{figure*}\centering
\subfigure[Evolution of neutrino masses]{\includegraphics[width=8cm]{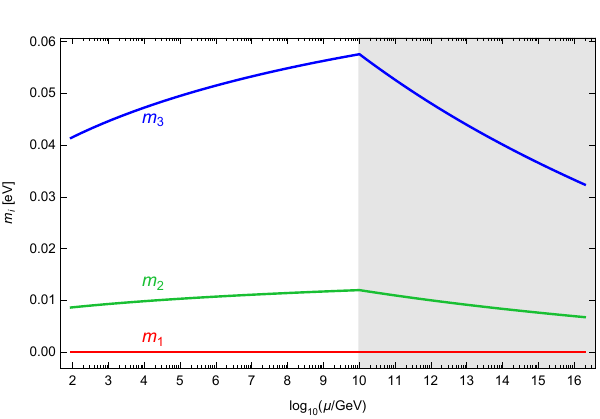}}
\subfigure[Evolution of the mass squared differences]{\includegraphics[width=8cm]{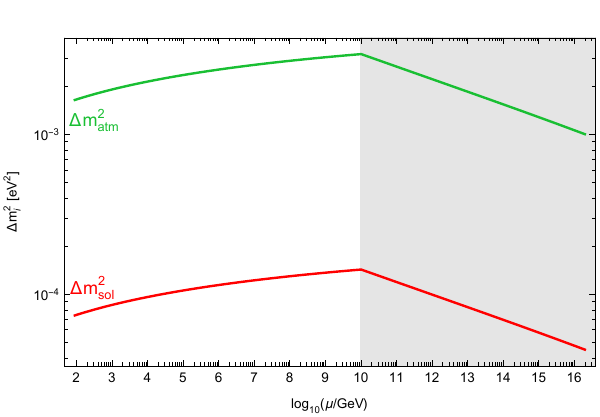}}
	\caption{As input values, we have chosen tribimaximal mixing at the GUT
  scale, $m_1=0\eV$, $\Delta m^2_\mathrm{atm}=10^{-3} \eV^2$,
  $\Delta m^2_\mathrm{sol}=4.5\times 10^{-5} \eV^2$, $M_\Delta=10^{10}\GeV$ and $\Lambda_6=2.5\times
  10^{-5} M_\Delta$, corresponding to $\braket{\Delta}=0.15\eV$. As we
  are only interested in showing the generic feature of the RG
  evolution, we choose the Higgs self-couplings to be 
  $\Lambda_{1,2,4,5}=0.5$ for simplicity, since they only indirectly influence the RG
  evolution of the angles and the flavor-dependent part of the RG
  equations of the masses.    
The shadowed area indicates the region where the Higgs triplet is
present. It is integrated out at the border between the shadowed and the
white area. \label{fig:Masses}}
\end{figure*}
Furthermore, the evolution of the mass squared difference is mainly given by 
\begin{equation}
16\pi^2 \frac{\Dot {\Delta m_{ji}^2}}{\Delta m_{ji}^2} \approx 2 \re \alpha + 4 C_\Delta \frac{m_j^2 + m_i^2}{v_\Delta^2} 
\end{equation}
in the SM and MSSM with small $\tan\beta$. There can be a cancellation
of the RG effect depending on the parameters $\Lambda_i$ in the Higgs
potential and the sign of $C_\Delta$, but generically the RG effect in
the effective theory is large, as it can be seen in
Fig.~\ref{fig:Masses}. This is just one possible example. The precise RG
effect strongly depends on the parameters in the Higgs potential
$\Lambda_i$.
The charged lepton Yukawa couplings depend on the singular values of Yukawa coupling matrix $Y_\Delta$ in a flavor non--diagonal way:
\begin{widetext}
\begin{subequations}
\begin{align}
16\pi^2 \frac{\Dot m_e}{m_e} = & \re \alpha_e + D_\Delta \left(\frac{m_1^2}{v_\Delta^2}\cos^2\theta_{12}  + \frac{m_2^2}{v_\Delta^2}\sin^2\theta_{12}\right) +\Ord{\theta_{13}}\\
16\pi^2 \frac{\Dot m_\mu}{m_\mu} = & \re \alpha_e + D_\Delta\left[ \frac{m_3^2}{v_\Delta^2} \sin^2\theta_{23}+\left(\frac{m_2^2}{v_\Delta^2}\cos^2\theta_{12}  + \frac{m_1^2}{v_\Delta^2}\sin^2\theta_{12} \right)\cos^2\theta_{23}\right] +\Ord{\theta_{13}}\\
16\pi^2 \frac{\Dot m_\tau}{m_\tau} = & \re \alpha_e + D_\Delta\left[ \frac{m_3^2}{v_\Delta^2}\cos^2\theta_{23} +\left( \frac{m_2^2}{v_\Delta^2} \cos^2\theta_{12}+\frac{m_1^2}{v_\Delta^2} \sin^2\theta_{12}\right)\sin^2\theta_{23}\right] + D_e y_\tau^2 +\Ord{\theta_{13}}\; .
\end{align}
\end{subequations}
\end{widetext}

\subsection{Running of the mixing angles}

Chao and Zhang~\cite{Chao:2006ye}  have derived the formulae in the approximation $|Y_e|\ll|Y_\Delta|$ which captures the dominant effects as long as there is a strong hierarchy. Here, we calculate the renormalization group equations exactly~\footnote{A Mathematica package with the exact formulae can be downloaded from \url{http://www.mpi-hd.mpg.de/\~mschmidt/rgeTriplet/}.}  and present the equations for the mixing angles in the approximation of vanishing $y_e,\,y_\mu$ and $\theta_{13}$:
\begin{widetext}
\begin{subequations}
\begin{align}
16\pi^2\Dot\theta_{12} = & -\frac{1}{2}\left[D_\Delta\frac{\Delta m_{21}^2}{v_\Delta^2}
+C_e y_\tau^2 \frac{\left|m_1\,e^{\I\varphi_1}+m_2\,e^{\I\varphi_2}\right|^2}{\Delta m_{21}^2}s_{23}^2\right]\sin2\theta_{12}+\Ord{\theta_{13}}\label{eq:thetaSol}\\
16\pi^2\Dot\theta_{13} = & -\frac{C_e}{2} y_\tau^2\frac{m_3\left[-m_1\Delta m_{32}^2 \cos\left(\delta-\varphi_1\right)+m_2\Delta m_{31}^2\cos\left(\delta-\varphi_2\right)+m_3 \Delta m_{21}^2\cos\delta\right]}{\Delta m_{31}^2\Delta m_{32}^2}
 \sin2\theta_{12}\sin2\theta_{23} +\Ord{\theta_{13}}\\
16\pi^2\Dot\theta_{23} = &-\frac{1}{2} \Bigg[D_\Delta \left(\frac{m_3^2}{v_\Delta^2}-\frac{m_1^2}{v_\Delta^2}s_{12}^2-\frac{m_2^2}{v_\Delta^2}c_{12}^2\right)
+C_e y_\tau^2
\left(\frac{\left|m_1\,e^{\I\varphi_1}+m_3\right|^2}{\Delta m_{31}^2}s_{12}^2
+\frac{\left|m_2\,e^{\I\varphi_2}+m_3\right|^2}{\Delta m_{32}^2}c_{12}^2\right)
\Bigg]\sin2\theta_{23}+\Ord{\theta_{13}}\; ,
\end{align}
\end{subequations}
\end{widetext}
where $C_e$ and $D_\Delta$ are defined in Eqs.~(\ref{eq:P},\ref{eq:F}) as well as $s_{ij}\equiv\sin\theta_{ij}$ and $c_{ij}\equiv\cos\theta_{ij}$.
The two contributions to the running from charged leptons and neutrinos
can be of the same order of magnitude and it strongly depends on the
hierarchy of neutrino masses which of the two contributions is dominant.
The contribution coming from the neutrino mass matrix ($\propto C_e$) shows almost the same features as in the effective theory:
\begin{itemize}
\item there is an enhancement factor which is proportional to $\frac{m_0^2}{\Delta m_{ji}^2}$, where $m_0$ denotes the mass scale of neutrinos;
\item the running strongly depends on $\tan\beta$ due to the charged lepton Yukawa couplings;
\item vanishing mixing is a fixed point.
\end{itemize}
On the other hand, the contribution from the charged leptons shows a completely different dependence on the Yukawa couplings. There is no dependence on the Majorana phases. It is basically proportional to the corresponding mass squared difference divided over the vev of the Higgs triplet squared. Hence, there is no large enhancement factor and no dependence on $\tan\beta$ in the SUSY case. Thus the overall size of the RG effect mainly depends on the vev of the Higgs triplet.
\begin{equation}
\Dot\theta_{ij}\sim\frac{\Delta m_{ji}^2}{v_\Delta^2}\sin2\theta_{ij}
\end{equation}
\begin{figure}[tb]
\centering
\includegraphics[width=8cm]{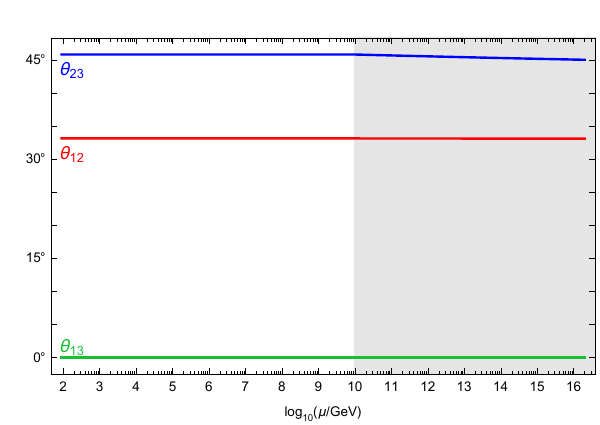}
\caption{Plot showing the evolution of the leptonic mixing angles in the
  SM. As input values, we have chosen tribimaximal mixing at the GUT
  scale, $m_1=0\eV$, $\Delta m^2_\mathrm{atm}=10^{-3} \eV^2$,
  $\Delta m^2_\mathrm{sol}=4.5\times 10^{-5} \eV^2$, $M_\Delta=10^{10}\GeV$ and $\Lambda_6=2.5\times
  10^{-5} M_\Delta$, corresponding to $\braket{\Delta}=0.15\eV$. As we
  are only interested in showing the generic feature of the RG
  evolution, we choose the Higgs self-couplings to be 
  $\Lambda_{1,2,4,5}=0.5$ for simplicity (In a realistic model, the
  parameters $\Lambda_i$ have to satisfy certain relations to produce
  the desired vevs.), since they only indirectly influence the RG
  evolution of the angles and the flavor-dependent part of the RG
  equations of the masses. In the MSSM with small $\tan\beta$, the
running is about twice as large because $C_\Delta$ is larger by a factor of two.
The shadowed area indicates the region where the Higgs triplet is
present. It is integrated out at the border between the shadowed and the
white area.\label{fig:Angles}}
\end{figure}
This gives a good estimate for the running in the strongly hierarchical case. The sign of the RG effect is determined by the sign of the mass squared difference and the factor $D_\Delta$ in front of the factor $Y_\Delta^\dagger Y_\Delta$ in $P$. As $D_\Delta$ is positive in the SM and MSSM, $\theta_{23}$ is evolving to larger values coming from the high renormalization scale for a normal hierarchy.
Furthermore, the $\beta$--function is approximately proportional to $\sin2\theta_{ij}$ which implies that a vanishing angle remains small. Taking into account these generic features, the RG effect from the charged leptons is largest for $\theta_{23}$ due to the combination of a large mass squared difference and a large angle.
As it can be seen from the equations, zero mixing is a fixed point. This
is also obvious from the RG equation in matrix form: in this configuration, $P$ and $F$ will be diagonal, if $Y_e$ and $Y_\nu$ are diagonal. 
In Fig.\ref{fig:Angles}, we have plotted the evolution of mixing angles in the SM for a strongly hierarchical spectrum in order to suppress the effect coming from the effective D5 operator. The gross features of the running can be immediately seen: the only sizable effect is on $\theta_{23}$ due to the large angle and mass squared difference. As it can be seen from the plot, the RG effect can be estimated by a leading log approximation to
\begin{equation}
\Delta\theta_{ij}\approx-\frac{D_\Delta}{2}\frac{\Delta m_{ji}^2}{v_\Delta^2}\sin2\theta_{ij}\ln\frac{\Lambda}{M_\Delta}\; .
\end{equation}
The contribution to $\theta_{13}$ coming from the charged leptons vanishes in our approximation. For non--vanishing $\theta_{13}$, it is given by
\begin{equation}
-\frac{D_\Delta}{2}\left(\frac{m_3^2}{v_\Delta^2}-\frac{m_1^2}{v_\Delta^2} \cos^2\theta_{12}-\frac{m_2^2}{v_\Delta^2}\sin^2\theta_{12}\right)\sin2\theta_{13}
\end{equation}
Let us comment on the configuration $\theta_{13}=m_3=0$, which is stable under the RG in the effective theory. 
Vanishing mass eigenvalues remain zero, as it can be seen from Eq. \eqref{eq:mDrei}, but $\theta_{13}$ receives corrections
\begin{equation}
\begin{split}
16\pi^2\Dot\theta_{13} = & \frac{D_\Delta}{2} \frac{\Delta m_{21}^2}{v_\Delta^2}\frac{y_e^2\left(y_\tau^2-y_\mu^2\right)}{\left(y_\tau^2-y_e^2\right)\left(y_\mu^2-y_e^2\right)} \cos\delta \sin2\theta_{12}\\
&\times\sin2\theta_{23} + \Ord{\theta_{13},\,y_3}
\end{split}
\end{equation}
Thus $\theta_{13}=m_3=0$ is not stable under the RG. However, the effect is negligible, because $\left(\frac{y_e}{y_\mu}\right)^2\frac{\Delta m_\mathrm{sol}^2}{v_\Delta^2}$ is very small and $m_3=0$ is stable.

\subsection{Running of the phases}
The RG evolution of the phases is described by:
\begin{widetext}
\begin{subequations}
\begin{align}
16\pi^2\Dot\delta = & 
C_e y_\tau^2 \Bigg[ \frac{1}{2\theta_{13}} \left(-\frac{m_1m_3}{\Delta m_{31}^2}\sin\left(\delta-\varphi_1\right)-\frac{m_2m_3}{\Delta m_{32}^2}\sin\left(\delta-\varphi_2\right)+\frac{m_3^2\Delta m_{21}^2}{\Delta m_{32}^2\Delta m_{31}^2}\sin\delta
\right)\sin2\theta_{12}\sin2\theta_{23}\\\nonumber
&+2\Bigg(
\frac{m_1m_2}{\Delta m_{21}^2}s_{23}^2\sin\left(\varphi_1-\varphi_2\right)
+\frac{m_1 m_3}{\Delta m_{31}^2}\cos\left(2\theta_{23}\right) s_{12}^2\sin\varphi_1
+\frac{m_2m_3}{\Delta m_{32}^2}\cos\left(2\theta_{23}\right)c^2_{12}\sin\varphi_2\\\nonumber
&+\frac{m_1m_3}{\Delta m_{31}^2}c^2_{23}c^2_{12}\sin\left(2\delta-\varphi_1\right)
+\frac{m_2m_3}{\Delta m_{32}^2}c^2_{23}s_{12}^2\sin\left(2\delta-\varphi_2\right)
\Bigg)
\Bigg] +\Ord{\theta_{13}}\\
16\pi^2\Dot\varphi_1 = & -4 C_e \left[\frac{m_1m_2}{\Delta m_{21}^2}s^2_{23}c^2_{12}\sin\left(\varphi_1-\varphi_2\right)
+\frac{m_1m_3}{\Delta m_{31}^2}\cos2\theta_{23}s^2_{12}\sin\varphi_1
+\frac{m_2m_3}{\Delta m_{32}^2}\cos2\theta_{23}c^2_{12}\sin\varphi_2
\right]
+\Ord{\theta_{13}^2}\\
16\pi^2\Dot\varphi_2 = &
-4 C_e \left[\frac{m_1m_2}{\Delta m_{21}^2}s^2_{23}s^2_{12}\sin\left(\varphi_1-\varphi_2\right)
+\frac{m_1m_3}{\Delta m_{31}^2}\cos2\theta_{23}s^2_{12}\sin\varphi_1
+\frac{m_2m_3}{\Delta m_{32}^2}\cos2\theta_{23}c^2_{12}\sin\varphi_2
\right]
+\Ord{\theta_{13}^2}\; .
\end{align}
\end{subequations}
\end{widetext}
Only the Dirac CP phase $\delta$ involves a term which is inversely proportional to $\theta_{13}$. Thus, there is a sizable effect for small $\theta_{13}$. For vanishing $\theta_{13}$, $\delta$ has to vanish (for realistic values of $\theta_{12}$ and $\theta_{23}$) in order to ensure analyticity of $\delta(t)$. 

\section{RG evolution in the full type II seesaw case\label{sec:fullMP}}

In the full type II case, it is not possible to express the RG equations in terms of mixing parameters. Therefore one has to resort to numerical calculations. For this purpose, we have extended the Mathematica package REAP, which is available on the web page \url{http://www.ph.tum.de/\~rge}, to include a left--handed triplet. 

\begin{figure}\centering
\subfigure[Angles]{\includegraphics[width=8cm]{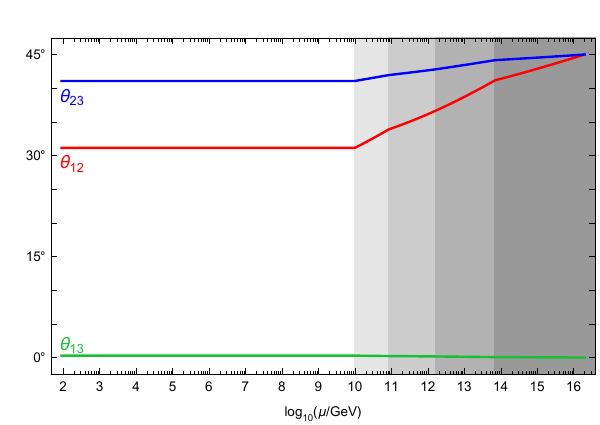}}
\subfigure[Masses]{\includegraphics[width=8cm]{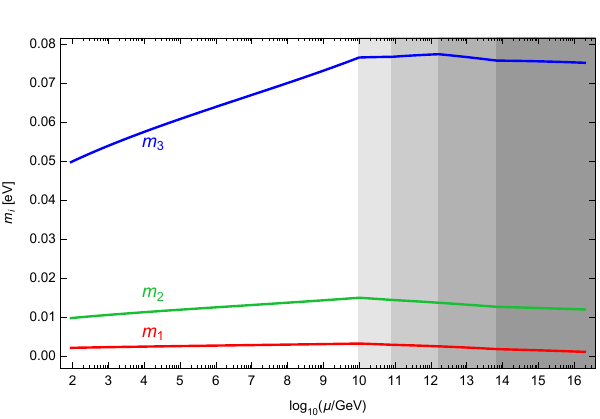}}
\caption{In the full type II seesaw case, there is a complicated
  interplay between the two contributions to the neutrino mass
  matrix. Here, we just plot an example for the following initial values
  at the GUT scale: $M_\Delta=10^{10}\GeV$, $\Lambda_{1,2,4,5}=0.1$, $\Lambda_6=8\times 10^9\GeV$,
  $m_1=10^{-3}\eV$, $\Delta m^2_\mathrm{sol}=1.4\times 10^{-4}\eV^2$,  $\Delta
  m^2_\mathrm{atm}=5.5\times 10^{-3}\eV^2$,
  $\theta_{12}=\theta_{23}=\frac{\pi}{4}$, $\theta_{13}=0$,
  $\delta=\varphi_1=\varphi_2=0$, $Y_\nu=0.4
  \diag\left(10^{-2},\;10^{-1},\;1\right)$, where $Y_\Delta$ is chosen
  diagonal $Y_\Delta=\diag\left(3.3\times
    10^{-7},\;3.9\times 10^{-6},\;2.5\times 10^{-5}\right)$ and $M$ is chosen
  appropriately to produce bimaximal mixing.
The differently shaded areas indicate the different energy ranges of the  
various effective field theories. At each border, a particle, either a
right-handed neutrino or the Higgs triplet, is integrated out.\label{fig:fullTypeII}}
\end{figure}

To illustrate the largeness of RG effects in the full type II seesaw
scenario, we show an example, where bimaximal mixing at high energy
evolves to the LMA solution at low energy. In previous
works~\cite{Antusch:2002hy,Miura:2003if,Shindou:2004tv,Antusch:2005gp},
this evolution was due to an inverted hierarchy in the neutrino Yukawa
couplings $Y_\nu$ or large imaginary off-diagonal entries. Here, the
relevant matrix $Y_\nu^\dagger Y_\nu$ is real and has a normal
hierarchy. In addition, the singular values of the Yukawa coupling
matrix $Y_\Delta$ are small ($\Ord{10^{-5}}$). In spite of the small couplings, there is a sizable effect on $\theta_{12}$ which can be seen in Fig.~\ref{fig:fullTypeII}. It is due to the different RG equations of the contributions to the neutrino mass matrix.

In our example, we have chosen $\Lambda_6$ to be relatively large $\Lambda_6=\Ord{10^9}\GeV$, because it receives corrections of the order of $M_3 \left(Y_\nu\right)_{33}^2 \left(Y_\Delta\right)_{33}$.
The evolution of the mixing angles $\theta_{12}$ and $\theta_{23}$ is non--linear above the threshold of the Higgs triplet and a leading log approximation is generally not possible. In the MSSM, the equations for the mixing angles presented in~\cite{Antusch:2005gp} are valid at each renormalization scale $\mu$. Hence, $\theta_{12}$ is increasing, as long as there are no imaginary off-diagonal entries and there is a normal hierarchy in the neutrino Yukawa couplings.

\section{Conclusions\label{sec:Conclusions}}

We calculated the RG equations in the type II seesaw case and found differences to the ones calculated by Chao and Zhang~\cite{Chao:2006ye} in the RG equations of the parameters of the Higgs potential. 

In the SM, the matrix $P$ describing the off--diagonal contributions to the RG evolution of the neutrino mass matrix is different for the contribution coming from the Higgs triplet compared to the one for the effective D5 operator. Hence, there can be large RG effects due to the different RG equations of the different contributions to the neutrino mass matrix.

Furthermore, we derived the exact RG equations in terms of the mixing parameters. The equations have a different structure compared to the ones in the standard seesaw case as well as in the effective theory case. 
The main difference to the running in the standard seesaw scenario is
the proportionality of the $\beta$-function of the mixing angles to the mass
squared difference in contrast to the inverse proportionality in the
case of a hierarchical spectrum. Hence, there is no enhancement factor
and the RG effect is small as long as $Y_\Delta$ is small. Furthermore,
the RG effect of the mixing angles $\theta_{ij}$ is proportional to $\sin2\theta_{ij}$ which leads together with the proportionality to the mass squared difference to a large effect on $\theta_{23}$ compared to the effect on $\theta_{12}$ due to $\Delta m^2_\mathrm{sol}\ll\Delta m^2_\mathrm{atm}$ and compared to $\theta_{13}$ due to $\sin2\theta_{13}\ll\sin2\theta_{23}$.

The RG equations in the full case can only be studied numerically. The interplay of the contributions from right--handed neutrinos and the Higgs triplet can lead to large RG effects even in the SM. Hence, it is necessary to consider RG effects in model building to make predictions which can be compared to the experimental data.

\section*{Acknowledgements}
M.S. wants to thank Werner Rodejohann for useful discussions and proofreading the manuscript, as well as He Zhang for pointing out an error in the calculation of the counterterms in the SM. The author also acknowledges support from the ``Deutsche Forschungsgemeinschaft'' in the ``Transregio Sonderforschungsbereich TR27: Neutrinos and Beyond'' and under project number RO--2516/3--2.

\section*{Note}
This updated version corrects two errors, which have been pointed out in \cite{Joaquim:2009vp}: 1) A wrong field normalization of the triplets has been used before, which amounts to a factor of two difference for all couplings of the triplets in the RG equations. 2) There was an error in the derivation of the RG equations of the mixing angles and phases. Both errors have been corrected in this version and the results agree with \cite{Joaquim:2009vp}.
MS would like to thank Guillem Domenech for pointing out two errors in the renormalization group equations~\cite{Domenech:2020yjf}. Two box diagrams with charged leptons have been missed for the counterterms of $\Lambda_{4,5}$ which lead to additional terms proportional to $\mathrm{tr}(Y_\Delta^\dagger Y_\Delta Y_e^\dagger Y_e)$ and the signs of the $\Lambda_{4,5}\Lambda_6$ terms in the counterterm and beta function of $\Lambda_6$ have to be reversed.
\appendix
\section{Counterterm Lagrangian}
\label{app:counterterm}

The relevant wave function renormalization factors are defined in the usual way by 
\begin{align}
    \left(\ell_L\right)_B=&Z_{\ell_L}^\frac{1}{2}\ell_L\\
  \Delta_B=&Z_\Delta^\frac{1}{2}\Delta\\
    \Higgs_B=&Z_\Higgs^\frac{1}{2}\Higgs\; .
\end{align}
The Yukawa couplings are renormalized multiplicatively
\begin{align}
    \left(\left(Z_{\ell_L}^T\right)^\frac{1}{2}\left(Y_\Delta\right)_B Z_{\ell_L}^\frac{1}{2}Z_\Delta^\frac{1}{2}\right)_{fg} =&  \mu^\frac{\epsilon}{2}\left(Y_\Delta  Z_{Y_\Delta}\right)_{fg}\\
    \left(Z_{\nu_R}^\frac{1}{2}\left(Y_\nu\right)_BZ_\Higgs^\frac{1}{2}Z_{\ell_L}^\frac{1}{2}\right)_{fg}=&\mu^\frac{\epsilon}{2}\left(Y_\nu Z_\nu\right)_{fg}\; .
\end{align}
The parameters in the Higgs potential have to be renormalized additively
\begin{subequations}
\begin{align}
    Z_\Higgs m_B^2=&m^2+\delta m^2\\
    Z_\Delta \left(M^2_\Delta\right)_B =&M_\Delta^2+\delta M_\Delta^2\\
    Z_\Higgs^2\lambda_B = & \mu^\epsilon Z_\lambda \lambda\\
    Z_\Delta^2\left(\Lambda_{1,2}\right)_B =& \mu^\epsilon \left(\Lambda_{1,2}+\delta\Lambda_{1,2}\right)\\
    Z_\Delta Z_\Higgs \left(\Lambda_{4,5}\right)_B =& \mu^\epsilon\left(\Lambda_{4,5}+\delta\Lambda_{4,5}\right)\\
    \left(Z^\dagger_\Delta\right)^\frac{1}{2}\left(Z_\Higgs^T\right)^\frac{1}{2}\left(\Lambda_6\right)_B
    Z_\Higgs^\frac{1}{2}=&\mu^\frac{\epsilon}{2}\left(\Lambda_6+\delta\Lambda_6\right)\; .
  \end{align}  
\end{subequations}
The insertion of the definitions of the renormalized quantities into the bare
Lagrangian yields the counterterm part of the Lagrangian which is needed to
cancel the divergences. Thus the counterterm Lagrangian
\begin{equation}
\mathcal{L}_\mathrm{ct}=\mathcal{C}_{\nu_R}+\mathcal{C}_\Delta
+\mathcal{C}_{V(\Delta)}
\end{equation}
is given by
\begin{widetext}
\begin{subequations}
  \begin{align}
   \mathcal{C}_{\nu_R}=&\frac{1}{2}\overline{\nu_R^g}\left(i\fmslash{\partial}\right)\left[\left(\delta Z_{\nu_R}\right)_{gf} \PL -\frac{1}{2}\overline{\nu_R}^g\left(\delta Z_M M\right)_{gf} \nu_R^f -\left(Y_\nu \delta Z_{Y_\nu}\right)_{gf} \overline{\nu_R}^g \tilde{\Higgs}^\dagger\ell_L^f + \left(\delta Z_{\nu_R}\right)_{gf}\PR\right]\nu_R^f +\hc\\
    \mathcal{C}_{\Delta}=&\delta Z_\Delta \tr\left(D_\mu\Delta\right)^\dagger \left(D^\mu\Delta\right)+ \left[i\tr\left(D_\mu\Delta\right)^\dagger\left(\delta Z_{g_1} g_1
    B^\mu\Delta+\delta Z_{g_2} g_2 \frac{\sigma_i}{2}\left[W^{i\mu},\Delta\right]\right)+\hc\right] \nonumber\\
&-\frac{Y_\Delta}{\sqrt{2}}\delta Y_\Delta
    \overline{\ChargeConjugate{\ell_L}}\left(i\sigma_2\right)\Delta\ell_L- \mathcal{C}_{V(\Delta)}\\
  \mathcal{C}_{V(\Delta)}=&\delta M_\Delta^2\tr\Delta^\dagger\Delta+\frac{\delta \Lambda_1}{2}\left(\tr\Delta^\dagger\Delta\right)^2
  +\frac{\delta\Lambda_2}{2}\left[\left(\tr\Delta^\dagger\Delta\right)^2-\tr\left(\Delta^\dagger\Delta\Delta^\dagger\Delta\right)\right]
  \nonumber\\
  &+\delta\Lambda_4\Higgs^\dagger\Higgs\tr\Delta^\dagger\Delta
  +\delta\Lambda_5\Higgs^\dagger\left[\Delta^\dagger,\Delta\right]\Higgs
  +\frac{\delta\Lambda_6}{\sqrt{2}}\Higgs^T\left(i\sigma_2\right)\Delta^\dagger\Higgs\; .
   \end{align}
\end{subequations}

\section{RG equations in the SM\label{app:RGE}}

The remaining RG equations of Yukawa coupling matrices  
\begin{align}
16\pi^2 \Dot Y_d  = &  Y_d \left[ \frac{3}{2} Y_d^\dagger Y_d -\frac{3}{2}\, Y_u^\dagger Y_u\right]+ Y_d\left[ T- \frac{1}{4} g_1^2 - \frac{9}{4} g_2^2 - 8\,g_3^2\right] \\
16\pi^2 \Dot Y_u  = & Y_u \left[ \frac{3}{2} Y_u^\dagger Y_u - \frac{3}{2}\, Y_d^\dagger Y_d\right]+Y_u\left[ T-\frac{17}{20} g_1^2 - \frac{9}{4} g_2^2 - 8\,g_3^2\right] ,
\end{align}
where $T=\Tr\left[Y_e^\dagger Y_e + Y_\nu^\dagger Y_\nu +3 Y_d^\dagger Y_d +3 Y_u^\dagger Y_u\right]$ and the right-handed neutrino mass matrix
\begin{equation}
16\pi^2 \Dot M =\left(Y_\nu Y_\nu^\dagger \right) M + M \left(Y_\nu Y_\nu^\dagger \right)^T \;.
\end{equation}
are listed for completeness. They are taken from ~\cite{Antusch:2005gp}.
The anomalous dimensions of the Higgs doublet mass is given by
\begin{equation}
  16\pi^2\gamma_m=\frac{9}{20}g_1^2+\frac{9}{4}g_2^2-\frac{3}{2}\lambda-T +\frac{2}{m^2}\tr\left(Y_\nu^\dagger M^2 Y_\nu\right)-3\Lambda_4\frac{M_\Delta^2}{m^2}-3\frac{|\Lambda_6|^2}{m^2}\; .
\end{equation}
and the renormalization of the remaining couplings in the Higgs potential are described by
	\begin{subequations}
  \begin{align}
  16\pi^2\Dot \lambda = &6 \lambda^2
  -3\lambda\left(3g_2^2+\frac{3}{5}g_1^2\right)+3g_2^4
	  + \frac{3}{2}\left(\frac{3}{5}g_1^2+g_2^2\right)^2
	  +4\lambda T\nonumber\\
&-8\tr\left(Y_e^\dagger Y_eY_e^\dagger Y_e+Y_\nu^\dagger Y_\nu Y_\nu^\dagger Y_\nu+3 Y_u^\dagger Y_u Y_u^\dagger
  Y_u+3 Y_d^\dagger Y_d Y_d^\dagger Y_d\right) +12\Lambda_4^2+8\Lambda_5^2\\
    16\pi^2\Dot \Lambda_1=&-\frac{36}{5}g_1^2\Lambda_1-24g_2^2\Lambda_1+\frac{108}{25}g_1^4\uline{+18g_2^4}+\frac{72}{5}g_1^2g_2^2\uline{+14\Lambda_1^2}+4\Lambda_1\Lambda_2+2\Lambda_2^2+4\Lambda_4^2+4\Lambda_5^2\nonumber\\
&+4\tr\left(Y_\Delta^\dagger Y_\Delta\right)\Lambda_1 \uline{-8\tr\left(Y_\Delta^\dagger Y_\Delta Y_\Delta^\dagger Y_\Delta\right)}\\
    16\pi^2\Dot \Lambda_2=&-\frac{36}{5}g_1^2\Lambda_2-24g_2^2\Lambda_2\uwave{+12g_2^4}-\frac{144}{5}g_1^2g_2^2+3\Lambda_2^2+12\Lambda_1\Lambda_2-8\Lambda_5^2
   +4\tr\left(Y_\Delta^\dagger Y_\Delta\right)\Lambda_2\nonumber\\
&+8\tr\left(Y_\Delta^\dagger Y_\Delta Y_\Delta^\dagger Y_\Delta\right)\\
  16\pi^2\Dot \Lambda_4=&-\frac{9}{2}g_1^2\Lambda_4-\frac{33}{2}g_2^2\Lambda_4+\frac{27}{25}g_1^4\uline{+6g_2^4}+\Bigg[8\Lambda_1+2\Lambda_2+3\lambda+4\Lambda_4+2 T + 2\tr\left(Y_\Delta^\dagger Y_\Delta\right)\Bigg]\Lambda_4\nonumber\\
  &+8\Lambda_5^2-4\tr\left(Y_\Delta^\dagger Y_\Delta Y_\nu^\dagger Y_\nu\right)
  -4\tr\left(Y_\Delta^\dagger Y_\Delta Y_e^\dagger Y_e\right)
  \\
  16\pi^2\Dot \Lambda_5=&-\frac{9}{2}g_1^2\Lambda_5-\frac{33}{2}g_2^2\Lambda_5-\frac{18}{5}g_1^2g_2^2+\Bigg[2\Lambda_1-2\Lambda_2+\lambda+8\Lambda_4\uline{+2 T + 2 \tr\left(Y_\Delta^\dagger Y_\Delta\right)}\Bigg]\Lambda_5\nonumber\\
&
+4\tr\left(Y_\Delta^\dagger Y_\Delta Y_\nu^\dagger Y_\nu\right)
-4\tr\left(Y_\Delta^\dagger Y_\Delta Y_e^\dagger Y_e\right)
\; .
\end{align}
\end{subequations}
There is also a contribution to the gauge coupling renormalization due to the Higgs triplet
\begin{align}
16\pi^2\beta_{g_1}=&\frac{41}{10}g_1^3+\frac{1}{6}\cdot 3\cdot 2 \frac{3}{5}g_1^3 = \frac{47}{10}g_1^3\\
16\pi^2\beta_{g_2}=&-\frac{19}{6}g_2^3+\frac{1}{6}\cdot 2 \cdot 2 g_2^3 = -\frac{5}{2}g_2^3\; ,
\end{align}
which has been also derived in ~\cite{Chao:2006ye}.
The RG equation of the strong coupling constant remains unchanged
\begin{equation}
16\pi^2\beta_{g_3}=-7 g_3^3\; .
\end{equation}

\section{RG equations in the MSSM\label{app:MSSM}}

The RG equations which do not directly influence the renormalization of the operator generating neutrino masses are given by:
\begin{align}
16\pi^2\Dot Y_e = &Y_e \left[ 3 Y_e^\dagger Y_e + Y_\nu^\dagger Y_\nu+3\, Y_\Delta^\dagger Y_\Delta\right] + Y_e\left[\tr(3 Y_d^\dagger Y_d + Y_e^\dagger Y_e)+3\,|\Lambda_d|^2-\frac{9}{5}g_1^2-3\,g_2^2\right]\\
16\pi^2\Dot Y_u = &Y_u \left[ Y_d^\dagger Y_d + 3 Y_u^\dagger Y_u\right] + Y_u\left[\tr(3 Y_u^\dagger Y_u + Y_\nu^\dagger Y_\nu)+3\,|\Lambda_u|^2-\frac{13}{15}g_1^2-3g_2^2-\frac{16}{3}g_3^2\right]\\
16\pi^2\Dot Y_d = &Y_d \left[3 Y_d^\dagger Y_d + Y_u^\dagger Y_u \right] + Y_d\left[\tr(3 Y_d^\dagger Y_d + Y_e^\dagger Y_e)+3\,|\Lambda_d|^2-\frac{7}{15}g_1^2-3g_2^2-\frac{16}{3}g_3^2\right]\\
16\pi^2\Dot\Lambda_d=&\Lambda_d\left[2\tr(3Y_d^\dagger Y_d+Y_e^\dagger Y_e)+7\,|\Lambda_d|^2+\tr(Y_\Delta^\dagger Y_\Delta)-\frac{9}{5}g_1^2-7g_2^2\right]
\end{align}
There are also contributions to the evolution of the gauge couplings of electroweak interactions:
\begin{align}
16\pi^2\beta_{g_1}=&\frac{33}{5}g_1^3+2\cdot 3 \frac{3}{5}g_1^3 = \frac{51}{5}g_1^3\\
16\pi^2\beta_{g_2}=&1g_2^3+2\cdot2\, g_2^3 = 5 g_2^3\; .
\end{align}
The RG equation of the strong coupling constant remains unchanged
\begin{equation}
16\pi^2\beta_{g_3}=-3 g_3^3\; .
\end{equation}
\end{widetext}

\renewcommand{\emph}[1]{{\it#1}}
\bibliography{rgeTriplet}
%Running,diplom,rge,ModelBuilding}
%\bibliographystyle{apsrev4}

\end{document}